\documentclass[lettersize,journal]{IEEEtran}
\usepackage{amsmath,amsfonts}
\usepackage{algorithmic}
\usepackage{algorithm}
\usepackage{array}
\usepackage{textcomp}
\usepackage{stfloats}

\usepackage{verbatim}
\usepackage{graphicx}
\usepackage{cite}
\usepackage[hyphens]{url}
\usepackage{breakurl}

\usepackage{cite}
\usepackage[english]{babel}
\usepackage{graphicx}%
\graphicspath{{./figures/}}
\usepackage{multirow}%
\usepackage{amsmath,amssymb,amsthm,amsfonts}
\usepackage{mathrsfs}%

\usepackage{xcolor}%
\usepackage{textcomp}%
\usepackage{manyfoot}%
\usepackage{booktabs}%
\usepackage{listings}%
\usepackage{tikz,lipsum,lmodern}

\usepackage{qcircuit}
\usepackage{amsmath,amssymb,amsthm,array}

\newtheorem{theorem}{Theorem}
\theoremstyle{definition}
\newtheorem{definition}[theorem]{Definition}
\usepackage{enumerate}

\usepackage{caption}
\usepackage{subcaption}

\usepackage{todonotes}

\newcommand{\ket}[1]{\ensuremath{\left|#1\right\rangle}} 
\newcommand{\bra}[1]{\ensuremath{\left\langle#1\right|}} 
\newcommand{\braket}[2]{\ensuremath{\left\langle#1|#2\right\rangle}} 

\renewcommand{\bf}[1]{\ensuremath{\mathbf{#1}}}

\begin{document}

\title{A Simple Quantum Blockmodeling with Qubits and  Permutations 
\author{Ammar Daskin
\IEEEauthorblockN{
                      \thanks{adaskin25@gmail.com}  
}
}}
\maketitle
\begin{abstract}
Blockmodeling of a given problem represented by an $N\times N$ adjacency matrix can be found by swapping rows and columns of the matrix (i.e. multiplying matrix from left and right by a permutation matrix).
Although classical matrix permutations can be  efficiently done by swapping pointers for the permuted rows (or columns) of the matrix,  by changing row-column order, a permutation changes the location of the matrix elements, which determines the membership of a group in the matrix based blockmodeling. Therefore, a brute force initial estimation of a fitness value for a candidate solution involving counting the memberships of the elements may require going through all the sum of the rows (or the columns).

Similarly permutations can be also implemented efficiently on quantum computers, e.g. a NOT gate on a qubit.
In this paper, using permutation matrices and qubit measurements, we show how to solve blockmodeling on quantum computers.
In the model, the measurement outcomes of a small group of  qubits are  mapped to indicate the fitness value. 
However, if the number of qubits in the considered group is much less than $n=log(N)$, it is possible to find or update the fitness value based on the state tomography in $O(poly(log(N)))$.
Therefore, when the number of iterations is less than $log(N)$ time and the size of the considered qubit group is  small, we show that it may be possible to reach the solution very efficiently.

\end{abstract}

\section{Introduction}

The research for quantum machine learning and data analysis started  almost as early as the quantum computing theory around the 1990s (see e.g. Ref. \cite{lewenstein1994quantum, kak1995quantum,  zak1998quantum,ezhov2000quantum, narayanan2000quantum} and  Ref.\cite{schuld2015introduction, cerezo2022challenges,jerbi2023quantum} for an introduction).
Because of the limitations in the number of controlled qubits, later studies, where quantum computing and machine learning are interleaved, are mostly related to quantum control where machine learning is used to better design (or control) qubits \cite{carrasquilla2020machine} or adiabatic quantum computation \cite{farhi2000quantum} where  a version of the adiabatic global optimization algorithm is used to design machine learning models \cite{albash2018adiabatic} and solve quadratic binary optimization problems \cite{date2021qubo, glover2018tutorial}.

Later machine learning approaches are generally based on some form of parameterized quantum circuits which can be broadly categorized into the following two different models: 
\begin{enumerate}[i)]
    \item For a given data in the forms of a group of vectors: $\bf{x_1}, \dots, \bf{x_N}$;  quantum machine learning models in general use the following parameterized equation \cite{jerbi2023quantum}:
\begin{equation}
    U(\bf{\theta})\ket{\bf{x_i}}
\end{equation}
Here, $\bf{\theta}$ is a vector of angle values that determine the parameters of the circuit. 
And \ket{\bf{x_i}} represents the encoded $i$th data which is either a quantum state vector that represents the vector of data vector $\bf{x_i}$ or one qubit used for each data point of the vector.
\item There are also models that use the following:
\begin{equation}
    U(\bf{\theta, x})\ket{\psi}
\end{equation}
Here, $\ket{\psi}$ is an initial state which has some simple implementation: i.e. in most cases an equal superposition or zero state, or a random separable state.
The difference from the generic variational circuit is that the data is no longer a quantum state and used to determine the angle values of the gates with the parameters (e.g. \cite{DaskinSimple2018,ghobadi2019power, perez2020data, yu2022power}).
\end{enumerate}
The latter models in most cases can be converted into the former one by using an extra ancillary quantum control register that uses the data \cite{jerbi2023quantum}. 
There is also an idea of the context aware quantum computation proposed in \cite{daskin2019context} which can be used to convert data entries into control qubits. 

Intuitions from quantum computing and algorithms tell us  quantum computers provide advantage when the superposition principle and faster sub-routines such as the quantum Fourier transform are fully employed.
In machine learning problems, the quantum Fourier transform can play the same role as in classical machine learning and provide a mechanism to do data analysis tasks in the frequency domain. 
Furthermore, in current variational quantum circuits, the superposition is believed to provide a better way to find the intricate relations in data. In certain models, it can also play a role to represent specific classical neural networks  more efficiently on quantum computers \cite{DaskinSimple2018}.

Although these machine learning models are able to mimic classical neural networks perfectly and some more accurate results specialized to certain problems are reported, to the best of our knowledge,   quantum machine learning models  proposed so far in the literature  are not proved to be more powerful or efficient in general theoretical comparison to classical machine learning when applied to classical data.
There is also a barren-plateaus problem in the training of huge dimensional data where the gradient value disappears (the optimization converges) before the model scans the whole solution space.
In addition, for successful learning, the number of required-parameters may be exponentially large.
Therefore, there is still a need to find a better way of modeling quantum machine learning where the advantage over classical machine learning is clear. 

We believe the superposition can parallelize data and decrease the training time drastically. 
In this paper, instead of considering data individually, we consider the blockmodeling problem as a single quantum state and show that by using different versions of the  permutation matrices, the solution to the blockmodeling problem can be obtained more efficiently in comparison to the similar classical approaches.
This paper is organized as follows: In the following sections, we first describe blockmodeling and show how qubit encoding and measurements on a group of qubits can be used to describe the blockmodeling. 
Then, we present different algorithms based on permutations and describe their classical and quantum implementations with their complexities.
In addition, we show numerical examples on the Barbell graph, discuss and give future directions in the applications related to clustering and classification and finally conclude the paper.

\section{Blockmodeling}
Graphs with random connection probabilities can be generated by following the Erdős–Rényi model where two nodes are connected with probability $p$. 
One can also generate the same model by following percolation theory and removing connections with probability $p$ from a fully connected graph. 
It is shown that not only the graphs representing some real world data but also these random models exhibit a high level of modularity. 
A module is defined as a highly connected subgraph with fewer connections with other parts (modules) of the graph. 
In networks, the modularity is used to measure the fraction of the edge assignments to an estimated community (a subgraph) which are better than a random assignment \cite{chodrow2021generative}.
Modularity of a graph can be also mapped into spin Hamiltonian where the ground state energy represents the modularity \cite{guimera2004modularity}.

Finding communities, components, clusters, cliques and many similar graph problems have applications in different fields of science and technology.
Blockmodeling  is used to remove perplexities  in  highly connected large networks and turn them into a comprehensible structure where meaningful relationships or clusters can be obtained easily \cite{doreian2005generalized,doreian2020advances,lee2019review} (These models are studied in different fields with different names. For unfamiliar readers we recommend Ref.\cite{abbe2017community} for an introduction.). 
There are many extensions and versions of blockmodeling or block clustering models \cite{kernighan1970efficient, lorrain1971structural,hartigan1972direct, white1976social, arabie1982blockmodels, faust1992blockmodels}: e.g. stochastic generalization of the model (i.e. stochastic block model)  used in machine learning and network science \cite{holland1983stochastic, anderson1992building,lee2019review}, Bayesian based models\cite{peixoto2019bayesian}, or hypergraph clustering \cite{chodrow2021generative}. 
Blockmodeling in general can be formulated as an optimization problem based on i) local optimizations (direct blockmodeling) by iteratively removing cluster members from one group to another or ii) a  matrix where dissimilarities among the considered clusters are maximized (indirect blockmodeling) \cite{Batagelj1992DirectIndirect, doreian1994partitioning, cugmas2019scientific}.

In terms of network analysis, a network is composed of units (actors, individuals) and links (relationships): In graph representation the units are vertices and links are the edges. 
The general idea of blockmodeling is to form two or more set of units into clusters where units in the same clusters are considered to have some form of  an equivalence such as structural equivalence where the connection of the units to the rest of the network are similar or regular equivalence where their connections with each other are similar \cite{batagelj1999generalized, batagelj2004generalized,doreian1999intuitive}. 

A blockmodel can be considered as a mapping of the units in a network onto the positions: In matrix formulation, this means finding a permutation of the rows and columns of the matrices in a way that the final structure of the matrix forms blocks. 
Based on the density and the connections to other blocks the blocks can be categorized, which are called building blocks. Therefore, the second step is to assign labels to the formed blocks based on the considered building blocks \cite{faust1992blockmodels}.
Note that there are also multi-layered versions of the problem \cite{valles2016multilayer}.

\section{ Quantum Version of the Problem Definition and Qubit Encoding of a Data Matrix}
Given similarity correlation matrix, or connection (adjacency) matrix; our goal is to find a sequence of permutation operations (row or column) that maps data vector in a state \ket{\bf{y}} so that $\braket{\bf{clusters}} {\bf{y}}  \approx 1$ or maximized. 
After this training, a cluster of data can be learned by measuring the group of qubits.
For this reason, we analyze qubit encoding and show how the states for the group of qubits can be used to measure the solution of a blockmodeling problem.

\subsection{Qubit encoding of a data matrix}
We consider a clustering problem (it could be similarly a classification problem in supervised learning or a pattern recognition problem), where we are given a data in the forms of a group of vectors: $\bf{x_1}, \dots, \bf{x_N}$. The task is to find similar data vectors and indicate which data points are in the same group (cluster or class).

In quantum computing, let us first convert data into a quantum state as follows: 
\begin{equation}
\ket{\bf{x}} = \left(\begin{matrix}
   \ket{\bf{x_1}}\\
   \vdots\\
   \ket{\bf{x_N}}
\end{matrix}\right)    
\end{equation}
The information about a quantum state is gained through the measurement of qubit states.
In general, a qubit state is defined with the vector elements where each half determines the zero and the one probabilites. 
If there is no pattern or structure in the data, we can expect to see almost equal probabilities in most cases. 
That means when a vector of uniformly random values is generated, a regular q-encoding will preserve the distribution in probabilities. 
Moreover, it is very likely to have  qubit-states defined with probabilities that are almost equal (an equal superposition state).
Although all qubit probabilities are determined with the equal number of states, the involved vector elements are different. 
Therefore, our goal is to use qubits to distinguish different clusters in the data. 

\begin{figure}[ht]
    \centering
    \includegraphics[width=0.95\linewidth]{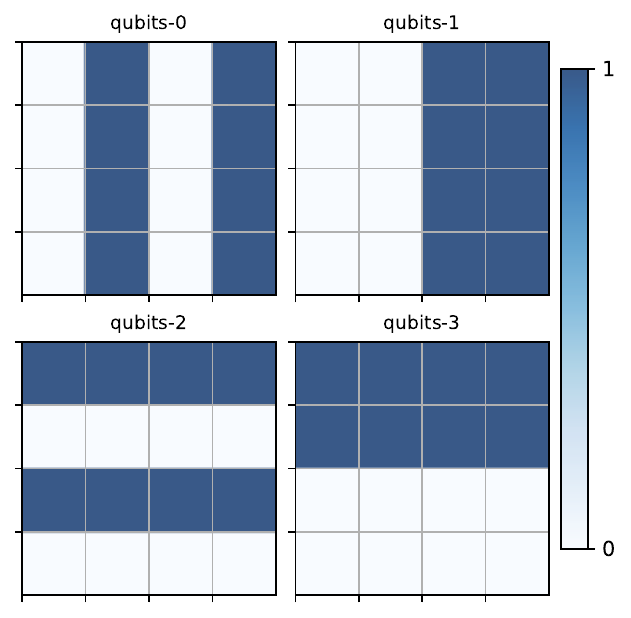}
    \caption{Qubit states in $16\times 16$ matrices: While blue represents $\ket{0}$, white represents   $\ket{1}$ states for the considered qubit. }
    \label{fig:cmap1}
\end{figure}

To see how qubit probabilities are determined and how different vector elements are involved in determining them; in Fig.\ref{fig:cmap1}, we show for a given matrix $X$ if we consider its vectorized form \ket{\bf{x}}, how different elements falls into the categories to determine the probability of a qubit.   
In the figures the dark color indicates the elements where a qubit is in \ket{0} state and the white color indicates where it is in \ket{1}. As seen from the figure, half of the elements indicate \ket{0} and the other half indicates \ket{1}.
If we have two groups, the probabilities of \ket{0} and \ket{1} for any chosen qubit can be used to measure the accuracy of blockmodeling for the problem vector \ket{\bf{x}}.

\begin{figure}[ht]
    \centering
    \includegraphics[width=0.95\linewidth]{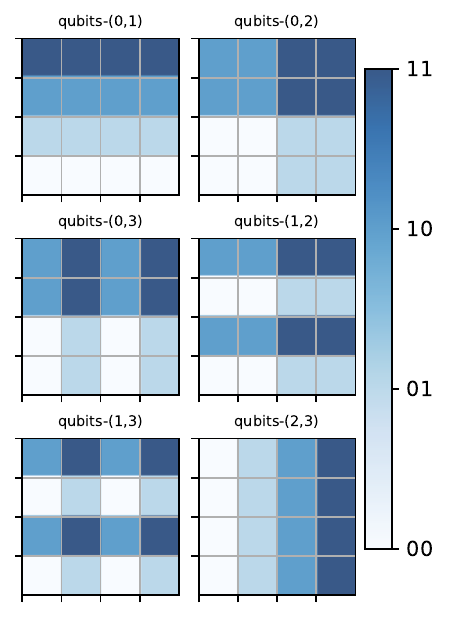}
    \caption{Group of two qubit-states in $4\times 4$ matrices. }
    \label{fig:cmap2}
\end{figure}

\begin{figure}[ht]
    \centering
    \includegraphics[width=0.95\linewidth]{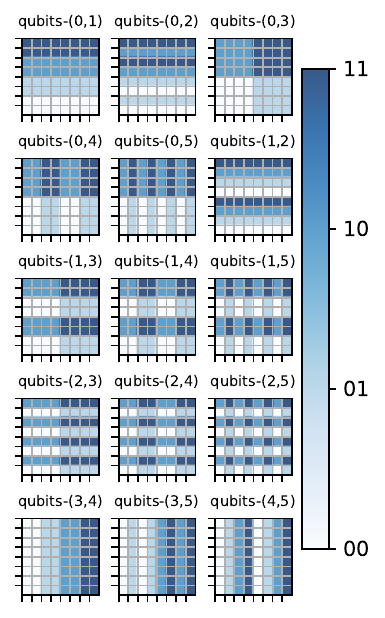}
    \caption{Group of two qubit-states in $8\times 8$ matrices.}
    \label{fig:cmap3}
\end{figure}

We can also consider a group of qubits together: In that case, for a group of two qubits, we have the states \ket{00}, \ket{01}, \ket{10}, and \ket{11} as shown in Fig.\ref{fig:cmap2} and Fig.\ref{fig:cmap3} for $4\times4$ and $8\times8$ data matrices (\ket{\bf{x}} is of dimensions 16 and 64, respectively.).

These figures show that the involved elements in determining the state of  a qubit or a group of qubits form a pattern in the matrix. 
We shall consider our machine learning model as overlapping the real clusters with these patterns so as to a qubit or a group of qubit to indicate the clusters.
And  the overlap-amount  will be the measurement of the model.

In many machine learning models, the data is converted into different forms and the number of dimensions are increased (e.g. kernel methods) or decreased (e.g. redundancy removal)
to make the model more efficient.
Here, in a general sense, the data is considered to be mapped by using a function $\Psi$.
Then, we solve a blockmodeling (or clustering) problem  by asking the following:
    Find a sequence of matrix permutation operations,$\prod_{i=1}^pP_i$, and a possible $\Psi$  so that the probabilities of at least a group of qubits in the following quantum state can distinguish the clusters in the data:
    \begin{equation}
        \ket{\bf{\hat{x}}} = \prod_{i=1}^pP_i\ket{\Psi(\bf{x})}.
    \end{equation}
$P_i$s only change the order of elements in a way that in the matrix form rows and columns are swapped. 
 Since we do not change matrix elements, and the model in general does not directly depend on a single amplitude, we can expect it to be more resilient to errors/noises from  measurements and gates. 
Furthermore, since the model is simple, it is easier to implement and do theoretical analyses which may be useful in certain applications. 

Here, note that one of the difficulties is that if the clusters are not equal to the number of elements, then we need to have dummies to swap them.
 Also, note that  we can  use a combination of many groups to decide the solution (similarly to random forest and boosting models). 
 In that case each of the subgroups of qubits can be considered a model.
In the training, if there are $k$  number of clusters, we will use all possible group of $k$ qubits for optimizing the parameters. 

As a result, the blockmodeling  problem can be formulated as:
\begin{definition}
      Finding a sequence of permutation operations that maps the data into the form so that a group of qubits represents the most clusters with best accuracy. 
\end{definition}
Note that each qubit can play a role in the decision with or without a grouping.
In addition, as in the decision trees, the number of possible groups can be as many as $O(n!)$. 
Therefore, based on the considered groups, the data space can be divided until reaching single data points.
This decision making and structure is  similar to the decision trees in classical machine learning.

\section{Algorithms Based on Permutations}
\label{Sec:Algorithms}
Blockmodeling can be solved by swapping the same column and row pairs: This way only the node orders change without changing their connections. 
For a problem in $N\times N$ matrix form of $\ket{\Psi(x)}$, if we choose indices  $a$ and $b$ so that $0\leq \{a,b\} \leq N$, we can define the following permutation that swaps $a$th and $b$th columns when applied from left and rows when applied from right to data matrix: 
\begin{equation}
\label{Eq:PabinGeneric}
    P_{ab} = \sum_{j = 0,j\neq a, b}^{N-1}  \ket{{j}} \bra{{j}} + \ket{{a}} \bra{{b}} + \ket{{b}} \bra{{a}},
\end{equation}
where $\ket{j}$ represents the $j$th vector in the standard basis. 

We also describe a fitness value for the candidate solution by first determining an expected probability vector, $\ket{expected}$ which represents expected blockmodeling. For instance, if we expect two blocks represented by 00, and 11 state, the expected probabilities are [05,0,0,0.5]. After obtaining quantum probabilities \ket{actual} for the considered two qubits, the fitness value of a candidate solution is defined as follows:
\begin{equation}
    \text{Fitness Value} =  \braket{expected}{actual}.
\end{equation}

Using the permutation in Eq.\eqref{Eq:PabinGeneric} along with the above defined fitness function, we can define a generic brute-force algorithm as follows:
\begin{enumerate}
    \item \label{stepab} At each iteration, choose $a$ and $b$, either random or based on some decision.
    \item \label{stepswap1}Swap columns $a$ and $b$
    \item \label{stepswap2}Swap rows $a$ and $b$
    \item Measure the fitness value of the solution
    \begin{enumerate}
        \item If there is an improvement over the best fitness, record the permutation and update the matrix and best fitness value
        \item Else disregard the permutation 
    \end{enumerate}
    \item If the stopping condition is satisfied or the maximum iteration number has been reached, then stop the iteration, otherwise continue on the next iteration.
\end{enumerate}
\subsection{Implementation and complexity on classical computers}
At each iteration, Step-\ref{stepab} of the algorithm requires $O(1)$ time. 
The permutations in Step-\label{stepwap1} and Step-\ref{stepswap2} similarly requires $O(1)$ time if we simply swap row and columns head pointers.
However, measuring and updating fitness value in Step-4 requires to go through all the vector elements in the $a$th and $b$th rows and columns. 
Therefore, in the worst case it requires $\Theta(N)$ complexity. 
Since there are $\Theta(N^2)$ possible $(a, b)$ pairs, in the worst case we need to repeat these steps $\Omega(N^2)$ times (Here also note that there are $O(N!)$ different number of permutations.). 
Therefore, in this worst case scenario, the complexity of these brute-force algorithms on classical computers is $\Omega(N^3)$. 
However, this lower bound is for the exact solution in the worst case. 
Considering this problem as an optimization problem, for the number of iterations $m$, we can describe the complexity of the optimization as $O(mN)$.

\subsection{Implementation and complexity on quantum computers}
\begin{figure}[ht]
    \input{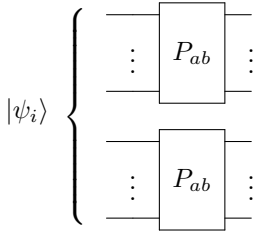}
    \caption{Generic quantum row and column swap operations applied to a quantum state at the $i$th iteration \ket{\psi_i}.}
    \label{fig:circuitPab} 
\end{figure}

\begin{figure}[ht]
    \input{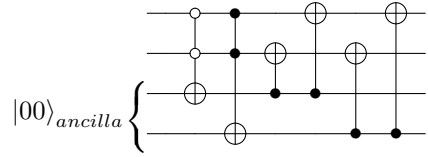}
    \caption{Implementation of swap $P_{ab}$ with the help of an ancillary register. Note that it can be implemented also without ancilla by a sequence of multi controlled swap operations. }
    \label{fig:circuitswap} 
\end{figure}

Consider \ket{\Psi(\bf{x})} and its qubit encoding similar to Fig.\ref{fig:cmap1}, Fig.\ref{fig:cmap2}, and Fig.\ref{fig:cmap3}. We can represent the matrix row and column indices with the register $\ket{\bf{r}}\ket{\bf{c}}$ with:
\begin{equation}
\begin{split}
\ket{\bf{r}} = \sum_{r=0}^N \ket{r}\\
\ket{\bf{c}} = \sum_{c=0}^N \ket{c}.
\end{split}
\end{equation}
Swapping the columns with the indices $a$ and $b$ in matrix form means swapping $\ket{\bf{r}}\ket{a}$ and $\ket{\bf{r}}\ket{b}$. 
Similarly for the columns, it means swapping 
$\ket{a}\ket{\bf{c}}$ and $\ket{b}\ket{\bf{c}}$. 
In other words,  we do the column-swap operations on the second register and the row-swap operations on the first register.
As a quantum circuit, we can represent these swap operations as $(P_{ab}\otimes P_{ab}) \ket{\psi_i}$ as shown in Fig.\ref{fig:circuitPab}. 
Here, $\ket{\psi_i}$ represents the quantum state at the $i$th iteration of the algorithm.
The implementation of $P{ab}$ requires multi-controlled $X$ (i.e. multi controlled CNOT) gates.
The implementation can be also done through using an ancillary register: 
Consider the quantum state \ket{00}+ \ket{01}+\ket{10}+\ket{11}. For $a=0$ and $b=3$, we can swap $\ket{00}$ and \ket{11} as in Fig.\ref{fig:circuitswap}.
Either $P_{ab}$ is implemented locally or by using an ancillary register, the number of quantum gates required for the implementation is $O(poly(n))$, considering $N = 2^n$.
The fitness value is found by considering the probability states of the group of qubits whose qubit encoding is considered for the solution of blockmodeling. 
If we consider the states of two qubits, measuring these qubits are enough to update the fitness value. 
Therefore, each iteration of the algorithm takes $O(poly(n))$ time. 
As in the classical case, in the worst case, for the exact solution, we need to repeat this algorithm $\Omega(N^2)$ times where we need to swap each column and rows at least once.
If the number of iterations is $m$, then we can describe the whole complexity as $\Theta(m\times poly(n))$, which could become  smaller than the classical complexity if $m\ll poly(n)$.

Note that the number of iterations can be decreased by implementing a group of permutations together, e.g. $P_{a_ib_i, a_jb_j}$ for some independent indices $\{a_i, b_i, a_j, b_j\}$, or by using $\left(P_{a_ib_i} \otimes P_{a_jb_j}\right)$ if they operate on different local terms, or by using a superpositioned ancilla register and swapping index pairs together (if the swap indices are independent). 
However, using an ancilla register may increase the complexity in the measurement because of the superpositioned ancilla adds ``extra parts" to the system and may decrease the success probability.

\subsection{Implementation of $P_{ab}$s in superposition}
As explained above, the action of $P_{ab}$ on matrix is to change columns $a$ and $b$: In terms of quantum state vector elements \ket{\bf{r}}\ket{a} and \ket{\bf{r}}\ket{b}. 
This means by using a select register similarly used in \cite{daskin2012universal,berry2015simulating},
we can apply a superposition of different $P_{ab}$s simultaneously.
However, we still need to find an efficient way to generate different choices of $P_{ab}$ by quantum operations.

Permutations on $\{0, 1, \dots, N-1\}$ form a non-Abelian group $S_N$ \cite{cameron1981finite
}. 
Any permutations $S_N$ can be written as a product of other permutations.
Using the cycle-notation for permutations; 
that means for instance, the permutation 
$\left(a_1 a_2 a_3 \dots  a_N\right)$ can be decomposed as $(a_1 a_N)(a_1 a_{N-1})\dots (a_1 a_3)(a_1 a_2)$, where $(ab)$ cyclically replaces $a$ with $b$ and $b$ with $a$. 

Another property of permutations is that if $(a\dots), (b\dots) \in S_N$ and disjoint cycles, then  the order, $r$ , of the product $(a\dots)(b\dots)$ is the least common multiple of the set lengths of $(a\dots)$ and $(b\dots)$. 
That means if we construct the permutation $P_{(a\dots)(b\dots)}$, the $r$th power of this matrix is the identity. And all the powers less than $r$ lead to another permutation matrix.


These observations shows that given an initial group of permutation matrices, we can define different permutations by applying $X$ gate on different combinations and by multiplying them together.
As an example, we can use the following construct:
\begin{itemize}
    \item Start with permutations that swaps neighbor elements: e.g. $P_{01}$ of dimension $2^n$ and e.g. $P_{12}$ of dimension $2^n$. 
    \item Construct the followings:
    \begin{equation}
    P_{0} = \left( I^{\otimes n} \right)\otimes P_{01} = 
    \left(\begin{matrix}
          P_{01}&&\\
          &\ddots&\\
          &&P_{01}\\
        \end{matrix}\right).
    \end{equation}
        \begin{equation}
    P_{1}=\left( I^{\otimes n}  \right)\otimes P_{12} = 
    \left(\begin{matrix}
          P_{12}&&&\\
          &\ddots&&\\
          &&&P_{12}\\
        \end{matrix}\right).
    \end{equation}
Here, $P_0$ and $P_1$ are of dimension $2^{2n}$. We can simply consider that we have an extra register \ket{select}. 

\item Define $\mathcal{X}$ controlled by a register \ket{select}:
\begin{equation}
\label{eq_matcalX}
\mathcal{X} = \bigotimes_{i=0}^n X^{k_i},
\end{equation} 
where $k_j$ is the $i$th bit in the binary of the indices encoded in the controlled register \ket{select}: i.e. if it has \ket{1} on the $j$th qubit, $X$ is applied to the $j$th qubit in the main register.
    
\item Apply $\mathcal{CX}$, i.e. $\mathcal{X}$ controlled by \ket{select}, to $P_{0}$ and $P_{1}$ to obtain the following (we assume \ket{select} in the equal superposition state.): 
\begin{equation}
\label{Eq:mathcalP0}
\mathcal{CX}.P_{0}=\left(\begin{matrix}
  P_{01}&&&\\
  &P_{23}&&&\\
  &&P_{45}&&\\
  &&&\ddots&\\
  &&&&P_{(N-2)(N-1)}\\
\end{matrix}\right)
\end{equation}
\begin{equation}
\small
\mathcal{CX}. P_{1}=\left(\begin{matrix}
  P_{12}&&&&&\\
  &P_{34}&&&&\\
  &&P_{56}&&&\\
  &&&\ddots&&\\
  &&&&&P_{(N-1)0}
  \\
\end{matrix}\right)
\end{equation}
We will call these operators  $\mathcal{P}_{0}$ and  $\mathcal{P}_{1}$.
As a circuit,  $\mathcal{P}_{0}$ is simply equivalent to Fig.\ref{fig:select}.
\begin{figure}
    \centering
    \input{fig_select}
    \caption{The circuit for  $\mathcal{P}_{0}$ defined in Eq.\eqref{Eq:mathcalP0}. }
    \label{fig:select}
\end{figure}
\end{itemize}
By starting with different permutations instead of $P_{01}$ and $P_{12}$ or by swapping permutations on the block diagonal of $\mathcal{P}_0$ and $\mathcal{P}_1$ using their product one can apply a superposition of permutations.

Measuring the output of the qubit group (the fitness value), in the collapsed state, \ket{select} would give the index that defines one of the permutations which give that fitness value.

\section{Numerical Example}

\subsection{A simple permuted Barbell graph}
\begin{figure*}[ht]
    \centering
     \begin{subfigure}[b]{1\linewidth}
     \centering
    \includegraphics[width=0.5\linewidth]{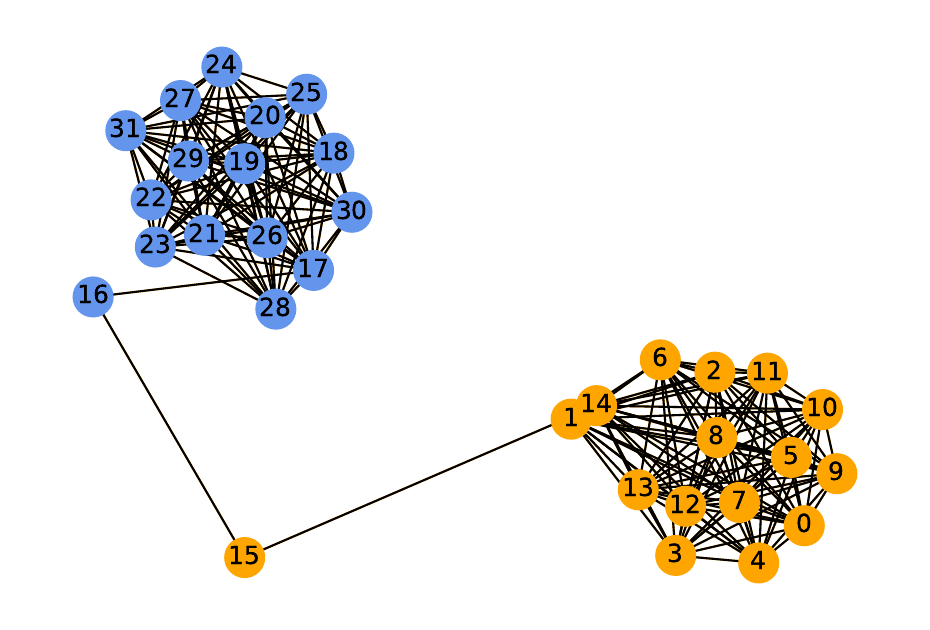}
    \caption{Graph representation}
    \label{fig:barbell-graph}
    \end{subfigure}
    \hfill
    \begin{subfigure}[b]{.49\linewidth}
    \centering
    \includegraphics[width=.8\linewidth]{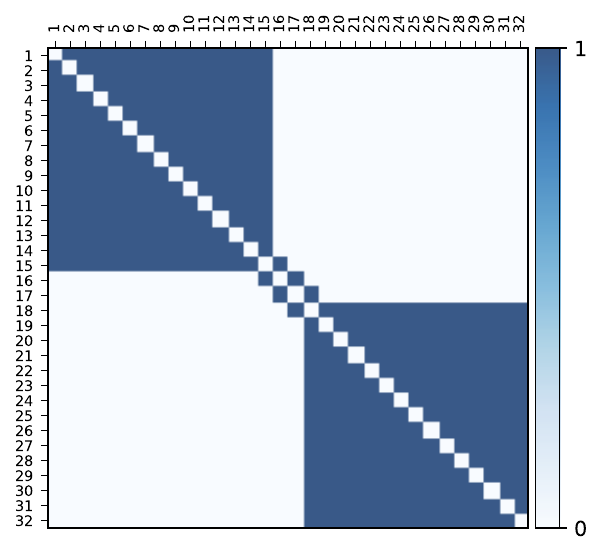}
    \caption{Matrix representation.}
    \label{fig:barbell-matrix}
    \end{subfigure}
    \begin{subfigure}[b]{.49\linewidth}
    \centering
    \includegraphics[width=.8\linewidth]{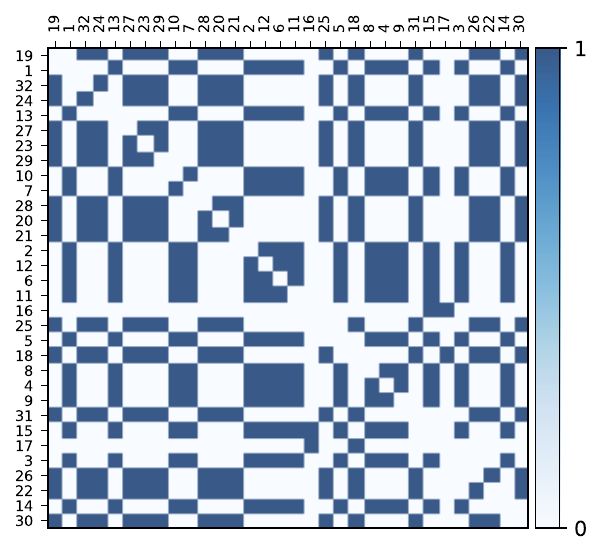}
    \caption{Shuffled matrix }
    \label{fig:barbell-shuffled}
    \end{subfigure}
    \caption{ (a) The graph representation of a 32 node Barbell graph. (b) Its matrix representation. Two qubit state-probabilities for the qubits $(0, 5)$ in this matrix are $[212,   1,   1, 212]$ without normalization  (c) The randomly shuffled problem matrix (The rows and columns of the original matrix is permuted) that is used in the example runs of the algorithm. 
    Two qubit state-probabilities for the qubits $(0, 5)$ for the shuffled matrix are $[114, 110, 110,  92]$ without normalization. }
    \label{fig:barbell}
\end{figure*}

In Fig.\ref{fig:barbell}, an example Barbell graph with 32 nodes is given. 
For illustration purposes, we will use the randomly permuted matrix in Fig.\ref{fig:barbell-shuffled} as the problem input and try to find the original matrix in Fig.\ref{fig:barbell-matrix}.

As qubit state encoding, since the matrix dimension is $32 \times 32$ which requires 10 qubits as a vector, we will use the states of 0th and 5th qubits in this 10 qubit system:  The state of the pair $(0, 5)$ gives a similar pattern to the pair $(0, 2)$ shown in Fig.\ref{fig:cmap2}.

The initial matrix in Fig.\ref{fig:barbell-matrix} is the goal for the optimization: i.e. We want to apply permutation matrices to convert the shuffled matrix to the original matrix so that the state probabilities for qubits (0, 5) becomes as close as possible to the probabilities in the original matrix, which is [212,   1,   1, 212] without normalization: Note that that these are the sum of matrix elements in each block represented by the states 00, 01, 10, 11 for the qubits 0 and 5. 
Furthermore, since there are empty rows and columns at the center of the matrix, there is more than one way to get this final state. 

\begin{figure*}[ht]
    \centering
     \begin{subfigure}[b]{0.49\linewidth}
     \centering
    \includegraphics[width=1\linewidth]{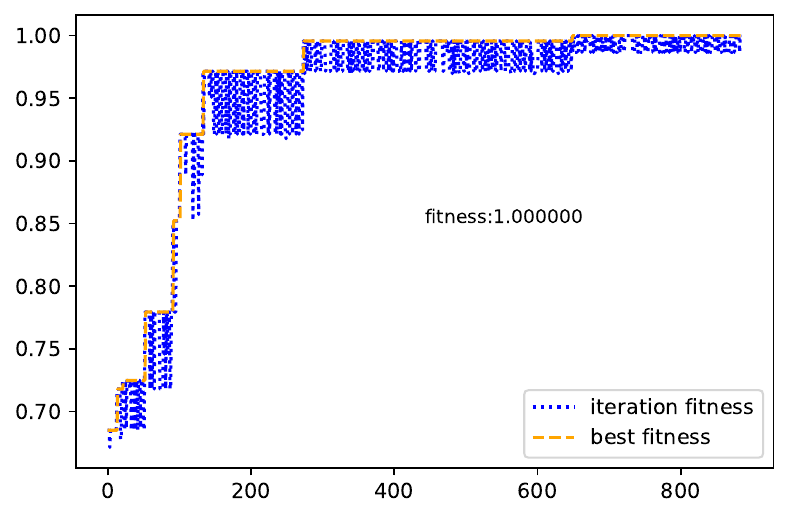}
    \caption{The fitness value during the iterations.}
    \label{fig:barbell-Pabruns}
    \end{subfigure}
    \begin{subfigure}[b]{0.49\linewidth}
    \centering
    \includegraphics[width=.8\linewidth]{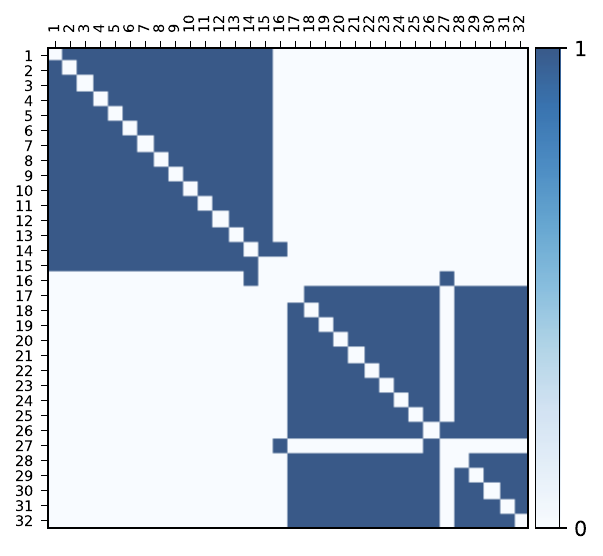}
    \caption{Matrix representation of the solution.
    }
    \label{fig:barbell-solutionmatrixPab}
    \end{subfigure}
    \caption{Solution for a Barbell graph problem given in Fig.\ref{fig:barbell-shuffled} by using random $P_{ab}$ gate. (a) The fitness value changes during the runs of the algorithm with random $P_{ab}$. (b) The final found solution matrix. 
    Two qubit state-probabilities for the qubits $(0, 5)$ are $[212,   1,   1, 212]$ without normalization. 
    Since in the original matrix, there is a gap in the centers, one row and column does not affect the probabilities for qubits (0, 5).}
    \label{fig:barbellrunswithPab}
\end{figure*}

We first use the circuit in Fig.\ref{fig:circuitPab} and follow the generic algorithm described at the beginning of Sec.\ref{Sec:Algorithms}: At each iteration of the runs, we generate $P_{ab}$ from random $a$ and $b$. Then update \ket{\psi_i}, if the fitness value is better than the best fitness value.
The best and observed fitness value at each iteration and the obtained solution matrix are shown in Fig.\ref{fig:barbellrunswithPab}.

\subsubsection{Restricting permutations for reducing the complexity and the number of iterations}
\begin{figure*}[ht]
    \centering
     \begin{subfigure}[b]{0.49\linewidth}
     \centering
    \includegraphics[width=1\linewidth]{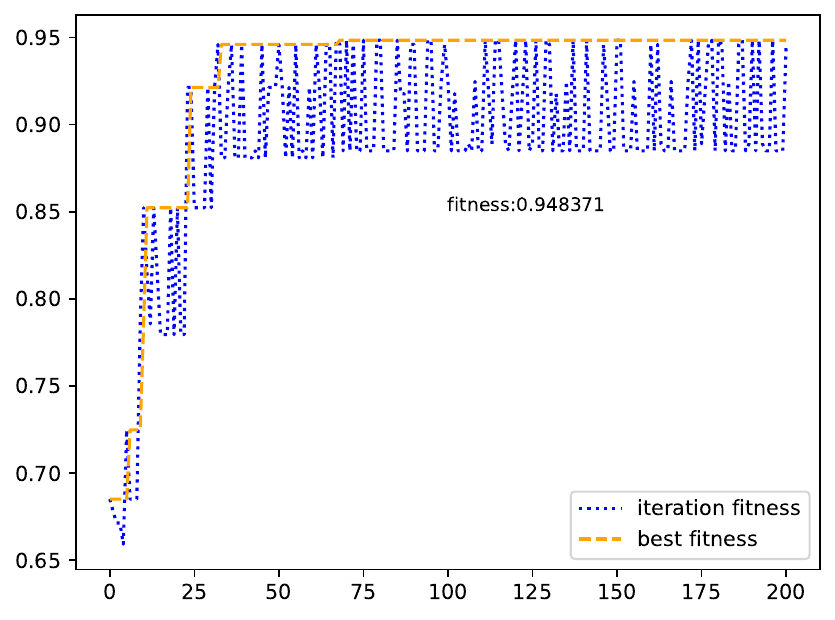}
    \caption{The fitness value during the iterations.}
    \label{fig:barbell-runs-MCX}
    \end{subfigure}
    \begin{subfigure}[b]{0.49\linewidth}
    \centering
    \includegraphics[width=.8\linewidth]{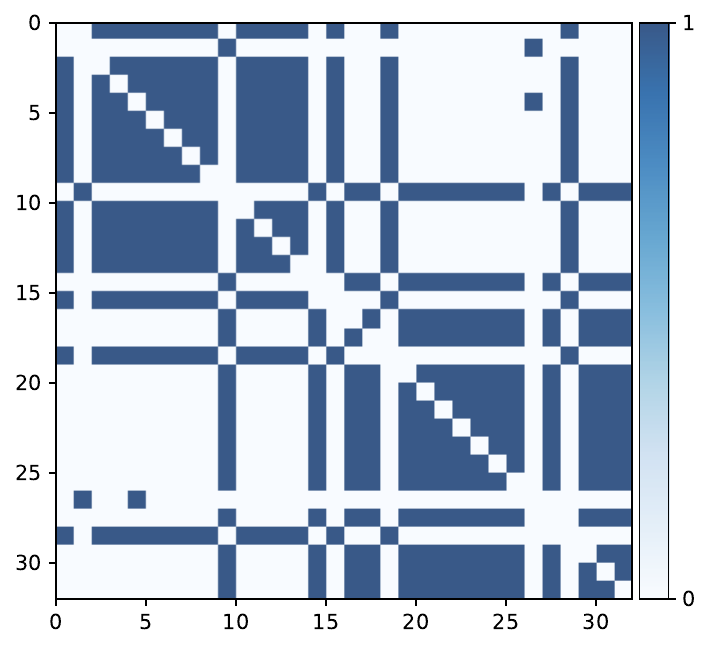}
    \caption{Matrix representation of the solution.
    }
    \label{fig:barbell-solutionmatrixMCX}
    \end{subfigure}
    \caption{Solution for a Barbell graph problem given in Fig.\ref{fig:barbell-shuffled} by using random $P_{ab}$ generated as a multi-controlled $X$ gate: The gate acts on the last qubits and controlled with the different states of the remaining qubits. 
    (a) The fitness value changes during the runs of the algorithm with random $P_{ab}$. (b) The final found solution matrix. 
    Two qubit state-probabilities for the qubits $(0, 5)$ are $[160,  54,  54, 158]$ without normalization. Because of the limitations enforced on the type permutations, this solutions is worse than the previous one. }
    \label{fig:barbellrunswithrandomMCX}
\end{figure*}

We can restrict permutations by using only a multi controlled gate that is controlled by different states of the first $n-1$ qubits and fix the target qubit to the last qubit. 
Because, in the matrix form of the quantum state, the blocks located on the diagonal are the ones we want to have 1 and on the reverse diagonal we want to have 0 entries. 
Therefore, we need to swap rows and columns crossing blocks on the  diagonal with the others.
In the index setting, \ket{r_0\dots r_{n-1}}\ket{c_0\dots c_{n-1}}, This can be done by setting target qubit in $P_{ab}$ for rows to ${r_{n-1}}$ and for columns to ${c_{n-1}}$  the last qubits.  
Fig.\ref{fig:barbellrunswithrandomMCX} shows the simulation runs of this setting: As can be seen from Fig.\ref{fig:barbell-runs-MCX}, the convergence to the solution is much faster: However, after converging a solution, it is not improving any more because of the limitations enforced on the type of the permutations.

\section{Discussion and Future Directions}
\subsection{The role of measured quantum state in optimization}
Since we know the parts of the states that need to be maximized, this information can be further used to design different optimization tools: For instance, one may integrate oblivious amplitude amplification which does not directly affect the amplitudes of the original system to increase the probabilities.

One can also use targeted permutations as we have done in the simulations by using multi controlled $X$ gate to reduce the number of iterations.

\subsection{New data samples and classification}
Classification methods are either based on distance based clustering(k-means) or direct clustering (decision trees)\cite{hartigan1972direct}.
In classification, a data set is given with $k$ class labels and "voting" based approaches can be used to determine the class of a new data sample.
There are well-studied quantum versions of k-means algorithms \cite{wiebe2015quantum,kerenidis2019q} which show how to load data from quantum RAM and output $k$-centroids that are the best representative of classes. 

The method described here can be formulated into classification or clustering problems as well by  using the qubit state to represent the mean value of the classes.
Since we know the labels and classes of data, one can also try reordering columns and rows based on a controlled register.
In this case, a distance base measurement may be used to determine the class label of a new data sample: i.e. either a quantum swap test or a quantum Hadamard test:
The swap test gives us a measure of the Euclidean distance of two vectors. It can be also incorporated in blockmodeling described here.

\subsection{Application to other problems}
Last but not least, it is known that the sampling in some image processing and computer vision task has circular matrix structures \cite{henriques2012exploiting}. 
It would be interesting to see whether those tasks also have sampling patterns that are similar to qubit encoding.

\section{Conclusion}
In this paper we have described how to do blockmodeling on quantum computers by using permutation matrices. We have explained how to build different permutations matrices and the complexity related issues which should be considered in their construction.
We have presented an example problem with a Barbell graph and its simulation results.
The permutations can be implemented efficiently on quantum computers and as explained in the paper an objective function of a blockmodeling or any similar problem can be related to the states of a group of a few qubits.
Since the states of a few qubits can be obtained efficiently, the quantum approach may provide  efficiency over the classical approaches where the fitness value of a candidate solution is found by summing all the vector elements.
\section{Data Availability}
All the graphs can be generated by using the code given in public repository on \url{https://github.com/adaskin/blockmodeling}.
\newpage
\bibliographystyle{IEEEtran}
\bibliography{main}

\begin{thebibliography}{10}
\providecommand{\url}[1]{#1}
\csname url@samestyle\endcsname
\providecommand{\newblock}{\relax}
\providecommand{\bibinfo}[2]{#2}
\providecommand{\BIBentrySTDinterwordspacing}{\spaceskip=0pt\relax}
\providecommand{\BIBentryALTinterwordstretchfactor}{4}
\providecommand{\BIBentryALTinterwordspacing}{\spaceskip=\fontdimen2\font plus
\BIBentryALTinterwordstretchfactor\fontdimen3\font minus \fontdimen4\font\relax}
\providecommand{\BIBforeignlanguage}[2]{{%
\expandafter\ifx\csname l@#1\endcsname\relax
\typeout{** WARNING: IEEEtran.bst: No hyphenation pattern has been}%
\typeout{** loaded for the language `#1'. Using the pattern for}%
\typeout{** the default language instead.}%
\else
\language=\csname l@#1\endcsname
\fi
#2}}
\providecommand{\BIBdecl}{\relax}
\BIBdecl

\bibitem{lewenstein1994quantum}
M.~Lewenstein, ``Quantum perceptrons,'' \emph{Journal of Modern Optics}, vol.~41, no.~12, pp. 2491--2501, 1994.

\bibitem{kak1995quantum}
S.~C. Kak, ``Quantum neural computing,'' \emph{Advances in imaging and electron physics}, vol.~94, pp. 259--313, 1995.

\bibitem{zak1998quantum}
M.~Zak and C.~P. Williams, ``Quantum neural nets,'' \emph{International journal of theoretical physics}, vol.~37, no.~2, pp. 651--684, 1998.

\bibitem{ezhov2000quantum}
A.~A. Ezhov and D.~Ventura, ``Quantum neural networks,'' \emph{Future Directions for Intelligent Systems and Information Sciences: The Future of Speech and Image Technologies, Brain Computers, WWW, and Bioinformatics}, pp. 213--235, 2000.

\bibitem{narayanan2000quantum}
A.~Narayanan and T.~Menneer, ``Quantum artificial neural network architectures and components,'' \emph{Information Sciences}, vol. 128, no. 3-4, pp. 231--255, 2000.

\bibitem{schuld2015introduction}
M.~Schuld, I.~Sinayskiy, and F.~Petruccione, ``An introduction to quantum machine learning,'' \emph{Contemporary Physics}, vol.~56, no.~2, pp. 172--185, 2015.

\bibitem{cerezo2022challenges}
M.~Cerezo, G.~Verdon, H.-Y. Huang, L.~Cincio, and P.~J. Coles, ``Challenges and opportunities in quantum machine learning,'' \emph{Nature Computational Science}, vol.~2, no.~9, pp. 567--576, 2022.

\bibitem{jerbi2023quantum}
S.~Jerbi, L.~J. Fiderer, H.~Poulsen~Nautrup, J.~M. K{\"u}bler, H.~J. Briegel, and V.~Dunjko, ``Quantum machine learning beyond kernel methods,'' \emph{Nature Communications}, vol.~14, no.~1, p. 517, 2023.

\bibitem{carrasquilla2020machine}
J.~Carrasquilla, ``Machine learning for quantum matter,'' \emph{Advances in Physics: X}, vol.~5, no.~1, p. 1797528, 2020.

\bibitem{farhi2000quantum}
E.~Farhi, J.~Goldstone, S.~Gutmann, and M.~Sipser, ``Quantum computation by adiabatic evolution,'' \emph{arXiv preprint quant-ph/0001106}, 2000.

\bibitem{albash2018adiabatic}
T.~Albash and D.~A. Lidar, ``Adiabatic quantum computation,'' \emph{Reviews of Modern Physics}, vol.~90, no.~1, p. 015002, 2018.

\bibitem{date2021qubo}
P.~Date, D.~Arthur, and L.~Pusey-Nazzaro, ``Qubo formulations for training machine learning models,'' \emph{Scientific reports}, vol.~11, no.~1, p. 10029, 2021.

\bibitem{glover2018tutorial}
F.~Glover, G.~Kochenberger, and Y.~Du, ``A tutorial on formulating and using qubo models,'' \emph{arXiv preprint arXiv:1811.11538}, 2018.

\bibitem{DaskinSimple2018}
A.~Daskin, ``A simple quantum neural net with a periodic activation function,'' in \emph{2018 IEEE International Conference on Systems, Man, and Cybernetics (SMC)}, 2018, pp. 2887--2891.

\bibitem{ghobadi2019power}
R.~Ghobadi, J.~S. Oberoi, and E.~Zahedinejhad, ``The power of one qubit in machine learning,'' \emph{arXiv preprint arXiv:1905.01390}, 2019.

\bibitem{perez2020data}
A.~P{\'e}rez-Salinas, A.~Cervera-Lierta, E.~Gil-Fuster, and J.~I. Latorre, ``Data re-uploading for a universal quantum classifier,'' \emph{Quantum}, vol.~4, p. 226, 2020.

\bibitem{yu2022power}
Z.~Yu, H.~Yao, M.~Li, and X.~Wang, ``Power and limitations of single-qubit native quantum neural networks,'' \emph{Advances in Neural Information Processing Systems}, vol.~35, pp. 27\,810--27\,823, 2022.

\bibitem{daskin2019context}
A.~Daskin, T.~Bian, R.~Xia, and S.~Kais, ``Context-aware quantum simulation of a matrix stored in quantum memory,'' \emph{Quantum Information Processing}, vol.~18, pp. 1--12, 2019.

\bibitem{chodrow2021generative}
P.~S. Chodrow, N.~Veldt, and A.~R. Benson, ``Generative hypergraph clustering: From blockmodels to modularity,'' \emph{Science Advances}, vol.~7, no.~28, p. eabh1303, 2021.

\bibitem{guimera2004modularity}
R.~Guimera, M.~Sales-Pardo, and L.~A.~N. Amaral, ``Modularity from fluctuations in random graphs and complex networks,'' \emph{Physical Review E}, vol.~70, no.~2, p. 025101, 2004.

\bibitem{doreian2005generalized}
P.~Doreian, V.~Batagelj, and A.~Ferligoj, \emph{Generalized blockmodeling}.\hskip 1em plus 0.5em minus 0.4em\relax Cambridge university press, 2005, no.~25.

\bibitem{doreian2020advances}
------, \emph{Advances in network clustering and blockmodeling}.\hskip 1em plus 0.5em minus 0.4em\relax John Wiley \& Sons, 2020.

\bibitem{lee2019review}
C.~Lee and D.~J. Wilkinson, ``A review of stochastic block models and extensions for graph clustering,'' \emph{Applied Network Science}, vol.~4, no.~1, pp. 1--50, 2019.

\bibitem{abbe2017community}
E.~Abbe, ``Community detection and stochastic block models: recent developments,'' \emph{The Journal of Machine Learning Research}, vol.~18, no.~1, pp. 6446--6531, 2017.

\bibitem{kernighan1970efficient}
B.~W. Kernighan and S.~Lin, ``An efficient heuristic procedure for partitioning graphs,'' \emph{The Bell system technical journal}, vol.~49, no.~2, pp. 291--307, 1970.

\bibitem{lorrain1971structural}
F.~Lorrain and H.~C. White, ``Structural equivalence of individuals in social networks,'' \emph{The Journal of mathematical sociology}, vol.~1, no.~1, pp. 49--80, 1971.

\bibitem{hartigan1972direct}
J.~A. Hartigan, ``Direct clustering of a data matrix,'' \emph{Journal of the american statistical association}, vol.~67, no. 337, pp. 123--129, 1972.

\bibitem{white1976social}
H.~C. White, S.~A. Boorman, and R.~L. Breiger, ``Social structure from multiple networks. i. blockmodels of roles and positions,'' \emph{American journal of sociology}, vol.~81, no.~4, pp. 730--780, 1976.

\bibitem{arabie1982blockmodels}
P.~Arabie, S.~A. Boorman \emph{et~al.}, ``Blockmodels: developments and prospects,'' in \emph{Classifying social data}.\hskip 1em plus 0.5em minus 0.4em\relax Jossey-Bass San Francisco, 1982, pp. 177--198.

\bibitem{faust1992blockmodels}
K.~Faust and S.~Wasserman, ``Blockmodels: Interpretation and evaluation,'' \emph{Social networks}, vol.~14, no. 1-2, pp. 5--61, 1992.

\bibitem{holland1983stochastic}
P.~W. Holland, K.~B. Laskey, and S.~Leinhardt, ``Stochastic blockmodels: First steps,'' \emph{Social networks}, vol.~5, no.~2, pp. 109--137, 1983.

\bibitem{anderson1992building}
C.~J. Anderson, S.~Wasserman, and K.~Faust, ``Building stochastic blockmodels,'' \emph{Social networks}, vol.~14, no. 1-2, pp. 137--161, 1992.

\bibitem{peixoto2019bayesian}
T.~P. Peixoto, ``Bayesian stochastic blockmodeling,'' \emph{Advances in network clustering and blockmodeling}, pp. 289--332, 2019.

\bibitem{Batagelj1992DirectIndirect}
V.~Batagelj, A.~Ferligoj, and P.~Doreian, ``Direct and indirect methods for structural equivalence,'' \emph{Social Networks}, vol.~14, no.~1, pp. 63--90, 1992, special Issue on Blockmodels.

\bibitem{doreian1994partitioning}
P.~Doreian, V.~Batagelj, and A.~Ferligoj, ``Partitioning networks based on generalized concepts of equivalence,'' \emph{Journal of Mathematical Sociology}, vol.~19, no.~1, pp. 1--27, 1994.

\bibitem{cugmas2019scientific}
M.~Cugmas, A.~Ferligoj, and L.~Kronegger, ``Scientific co-authorship networks,'' \emph{Advances in network clustering and blockmodeling}, pp. 363--387, 2019.

\bibitem{batagelj1999generalized}
V.~Batagelj, A.~Ferligoj, and P.~Doreian, ``Generalized blockmodeling,'' \emph{Informatica(Ljubljana)}, vol.~23, no.~4, pp. 501--506, 1999.

\bibitem{batagelj2004generalized}
V.~Batagelj, A.~Mrvar, A.~Ferligoj, and P.~Doreian, ``Generalized blockmodeling with pajek,'' \emph{Advances in methodology and statistics}, vol.~1, no.~2, pp. 455--467, 2004.

\bibitem{doreian1999intuitive}
P.~Doreian, ``An intuitive introduction to blockmodeling with examples,'' pp. 5--34, 1999.

\bibitem{valles2016multilayer}
T.~Valles-Catala, F.~A. Massucci, R.~Guimera, and M.~Sales-Pardo, ``Multilayer stochastic block models reveal the multilayer structure of complex networks,'' \emph{Physical Review X}, vol.~6, no.~1, p. 011036, 2016.

\bibitem{daskin2012universal}
A.~Daskin, A.~Grama, G.~Kollias, and S.~Kais, ``Universal programmable quantum circuit schemes to emulate an operator,'' \emph{The Journal of chemical physics}, vol. 137, no.~23, 2012.

\bibitem{berry2015simulating}
D.~W. Berry, A.~M. Childs, R.~Cleve, R.~Kothari, and R.~D. Somma, ``Simulating hamiltonian dynamics with a truncated taylor series,'' \emph{Physical review letters}, vol. 114, no.~9, p. 090502, 2015.

\bibitem{cameron1981finite}
P.~J. Cameron, ``Finite permutation groups and finite simple groups,'' \emph{Bulletin of the London Mathematical Society}, vol.~13, no.~1, pp. 1--22, 1981.

\bibitem{wiebe2015quantum}
N.~Wiebe, A.~Kapoor, and K.~M. Svore, ``Quantum algorithms for nearest-neighbor methods for supervised and unsupervised learning,'' \emph{Quantum Information \& Computation}, vol.~15, no. 3-4, pp. 316--356, 2015.

\bibitem{kerenidis2019q}
I.~Kerenidis, J.~Landman, A.~Luongo, and A.~Prakash, ``q-means: A quantum algorithm for unsupervised machine learning,'' \emph{Advances in neural information processing systems}, vol.~32, 2019.

\bibitem{henriques2012exploiting}
J.~F. Henriques, R.~Caseiro, P.~Martins, and J.~Batista, ``Exploiting the circulant structure of tracking-by-detection with kernels,'' in \emph{Computer Vision--ECCV 2012: 12th European Conference on Computer Vision, Florence, Italy, October 7-13, 2012, Proceedings, Part IV 12}.\hskip 1em plus 0.5em minus 0.4em\relax Springer, 2012, pp. 702--715.

\end{thebibliography}
\end{document}